\documentclass{article}
\usepackage{arxiv}
\usepackage[utf8]{inputenc}
\usepackage[T1]{fontenc}
\usepackage{hyperref}
\usepackage{url}
\usepackage{booktabs}
\usepackage{amsfonts}
\usepackage{nicefrac}
\usepackage{microtype}
\usepackage{graphicx}
\usepackage{tabularx}
\usepackage{multirow}
\usepackage{xcolor}

\title{FIST: A Structured Threat Modeling Framework for Fraud Incidents}

\author{
  Yu-Chen Dai \thanks{National Institute of Cyber Security, Tainan, Taiwan. \texttt{yuchen.dai@nics.nat.gov.tw}} \\
  \And
  Lu-An Chen \thanks{National Institute of Cyber Security, Tainan, Taiwan. \texttt{luan.cs11@nycu.edu.tw}} \\
  \And
  Sy-Jye Her \thanks{National Institute of Cyber Security, Taipei, Taiwan. \texttt{jessieher@nics.nat.gov.tw}} \\
  \And
  Yu-Xian Jiang \thanks{National Institute of Cyber Security, Tainan, Taiwan. \texttt{yuxianjiang@nics.nat.gov.tw}} \\
}

\begin{document}
\maketitle

\begin{abstract}
Fraudulent activities are rapidly evolving, employing increasingly diverse and sophisticated methods that pose serious threats to individuals, organizations, and society. This paper proposes the FIST Framework (Fraud Incident Structured Threat Framework), an innovative structured threat modeling methodology specifically designed for fraud scenarios. Inspired by MITRE ATT\&CK and DISARM, FIST systematically incorporates social engineering tactics, stage-based behavioral decomposition, and detailed attack technique mapping into a reusable knowledge base. FIST aims to enhance the efficiency of fraud detection and the standardization of threat intelligence sharing, promoting collaboration and a unified language across organizations and sectors. The framework integrates interdisciplinary insights from cybersecurity, criminology, and behavioral science, addressing both technical vectors and psychological manipulation mechanisms in fraud. This approach enables fine-grained analysis of fraud incidents, supporting automated detection, quantitative risk assessment, and standardized incident reporting. The effectiveness of the framework is further validated through real-world case studies, demonstrating its value in bridging academic research and practical applications, and laying the foundation for an intelligence-driven anti-fraud ecosystem. To the best of our knowledge, FIST is the first systematic, open-source fraud threat modeling framework that unifies both technical and psychological aspects, and is made freely available to foster collaboration between academia and industry.
\end{abstract}




\section{Introduction}

In recent years, the widespread adoption of digital communication and social platforms has driven the rapid evolution of fraud techniques, with victims no longer limited to individuals but extending to the stability of financial markets and the credibility of government sectors. Compared to traditional cybersecurity threats, fraud increasingly relies on advanced social engineering techniques—such as impersonating government agencies, customer service representatives, or financial advisors—and even leverages AI-powered deepfake technologies. Although existing frameworks like MITRE ATT\&CK provide important references for threat modeling, they remain insufficient for fully capturing the unique social engineering behaviors and multi-stage tactics characteristic of real-world fraud, whose attack complexity now rivals that of major cybersecurity threats.


Recent studies have emphasized the urgent need for structured and interoperable approaches to analyze, track, and prevent cybercrime and fraud. Sarkar et al.\cite{sarkar2023} proposed a TTPs (Tactics, Techniques, and Procedures)-based cybercrime investigation framework, covering 14 tactics, 177 techniques, and 303 sub-techniques, and demonstrated its effectiveness through case studies. The IIT Kanpur team further developed investigation tools that operationalize these frameworks, enabling evidence tracking and attack path analysis\cite{iitkanpur2023}. Meanwhile, Alhashmi et al.~\cite{alhashmi2024} conducted a comprehensive analysis of vulnerabilities in online fraud, integrating psychological, behavioral, technical, and environmental factors to propose prevention and mitigation strategies.


Despite these advances, most existing frameworks remain focused on technical vulnerabilities or are limited to high-level descriptions, making it difficult to fully capture the hybrid nature of modern fraud within a modular and extensible architecture. Modern fraud often involves dynamic, multi-stage social engineering and adaptive psychological manipulation. While some domain-specific fraud knowledge bases exist, they do not provide the extensible, phase-based modeling and systematic mapping capabilities found in ATT\&CK, DISARM, or the FIST framework proposed in this study. To address this gap, the FIST framework is designed as an open, extensible, and empirically validated solution, released as an open-source project and continuously maintained by the community. FIST is committed to promoting transparency and a spirit of collaboration, accelerating academic research, practical adoption, and rapid adaptation to emerging fraud threats. It supports automated detection, collaborative intelligence sharing, and ongoing optimization, driving the development of intelligent anti-fraud solutions that effectively meet the needs of real-world scenarios.


For the benefit of the research and practitioner communities, the framework (including documentation and case studies) will be made publicly available as open-source after publication.


\section{FIST Framework Overview}

The FIST framework draws inspiration from MITRE ATT\&CK~\cite{mitre_attack}, DISARM~\cite{disarm}, and the structured modeling paradigm of STIX~\cite{stix}, and is purpose-built to address the unique challenges of fraud. Unlike these prior frameworks, which primarily focus on cyber threats or disinformation operations, FIST integrates both the technical attack vectors and psychological manipulation strategies commonly found in modern fraud—such as inducing urgency and fear, exploiting intimate relationships, and leveraging authority pressure. The framework decomposes the fraud lifecycle into distinct operational phases, with a focus on the attacker’s objectives, assets, and multi-layered tactics. This approach facilitates modular expansion and case-based application, promoting cross-sector knowledge sharing and local adaptation.


\subsection{Design Rationale and Structure}

Each phase of the FIST framework corresponds to specific attacker mindsets, operational objectives, and a distinct set of observable indicators. By explicitly mapping various tactics and techniques—such as SMS bombing, AI-powered fake persona creation, and multi-platform message dissemination—FIST enables precise tracking of attack paths and identification of detection opportunities. This process-centric perspective not only supports forensic analysis, but also guides defenders in proactively disrupting key nodes within the fraud kill chain.


Compared to recent TTP-based cybercrime frameworks~\cite{sarkar2023}—which, for example, catalog 14 tactics, 177 techniques, and over 300 sub-techniques—FIST offers a streamlined but equally systematic approach, tailored to the fraud ecosystem. The current FIST release comprises 4 operational phases, 9 major tactics, and 93 detailed techniques (including sub-techniques), with 58 detection patterns, 12 mitigations, and 12 tool entries, all structured for modular extensibility (see Table~\ref{tab:fist_scale}).


\begin{table}[htbp]
\caption{FIST Framework Components and Scale}
\centering
\begin{tabular}{lc}
\toprule
\textbf{Component} & \textbf{Count} \\
\midrule
Operational Phases & 4 \\
Major Tactics & 9 \\
Detailed Techniques & 93 \\
Detection Patterns & 58 \\
Mitigation Strategies & 12 \\
Tool Entries & 12 \\
\bottomrule
\end{tabular}
\label{tab:fist_scale}
\end{table}

FIST organizes fraud incidents into four major phases:
\begin{enumerate}
\item \textbf{Preparation}: Reconnaissance and creation of deceptive assets (e.g., phishing sites, fake identities).
\item \textbf{Promotion}: Dissemination of fraudulent messages and luring of victims via psychological and contextual triggers.
\item \textbf{Engagement}: Building trust and guiding victims toward harmful actions (e.g., fund transfer, disclosure).
\item \textbf{Concealment}: Covering tracks, laundering gains, and removing evidence.
\end{enumerate}


The FIST framework adopts a hierarchical design, with each phase comprising high-level tactics further broken down into specific techniques. This layered structure enables fine-grained analysis, systematic comparison, and effective knowledge sharing across organizations. Unlike conventional cybersecurity frameworks, FIST is unique in its dual-track approach, simultaneously covering both psychological manipulation and technical attack vectors, making it highly effective for addressing today’s hybrid fraud scenarios—such as those that combine social engineering with AI-generated deepfake technologies.


Leveraging its modular architecture and standardized taxonomy, FIST facilitates seamless integration with automated platforms and threat intelligence systems. By applying FIST-based tagging and classification to incidents and indicators, organizations can achieve consistent incident reporting, streamline knowledge transfer, and enable collaborative defense mechanisms. This approach not only enhances interoperability between security solutions and intelligence protocols, but also supports industry-wide cooperation for more effective and scalable anti-fraud operations.


\section{Case Study: Investment Fraud via Social Media}

We demonstrate the application of FIST through a detailed analysis of a real-world investment fraud operation on a messaging platform.


\subsection{Incident Summary}

Victims were enticed to join a group by individuals posing as financial experts who promised high returns. The fraud scheme involved personalized investment advice, staged trust-building, and gradually increasing financial requests. The fraudsters manipulated the withdrawal process and used fake testimonials or time-limited offers to intensify psychological pressure and evade detection. This case also demonstrated dynamic adaptation: when victims grew suspicious or hesitated, the scammers deployed additional psychological manipulation tactics—such as presenting fabricated testimonials or creating limited-time investment windows—to increase pressure. Such adaptive strategies can be systematically tracked using FIST, offering valuable insights for defender playbooks and automated alert tuning.


\subsection{FIST Technique Mapping and Detection Points}

\begin{table}[htbp]
\caption{Investment Fraud Case Study: FIST Technique Mapping and Detection Points}
\centering
\renewcommand{\arraystretch}{1.7}
\setlength{\tabcolsep}{2pt}
\small
\begin{tabularx}{\textwidth}{|p{2.3cm}|p{2.9cm}|X|X|}
\hline
\textbf{FIST Phase} & \textbf{Technique ID \& Name} & \textbf{Observed Behavior} & \textbf{Detection Indicators} \\
\hline

\multirow{4}{2.3cm}{Preparation (P0001)} 
& T0003: Social Media Analysis & Analyzed victim profiles to identify investment interests and financial capacity & D0002.001: Abnormal account activity patterns, rapid profile browsing \\
\cline{2-4}
& T0009.002: Fake Investment Text Creation & Created high-return investment plans promising ``30\% monthly profits'' & D0001.010: Fraud keywords detection (``guaranteed profit'', ``risk-free'') \\
\cline{2-4}
& T0010.001: Fake Account Creation & Established professional investment advisor personas with AI-generated avatars & D0001.001: AI-generated content detection via deepfake analysis \\
\cline{2-4}
& T0012: Fraudulent Website Creation & Built fake investment platforms mimicking legitimate brokers & D0003.001: Suspicious domain registration, typosquatting \\
\hline

\multirow{3}{2.3cm}{Promotion (P0002)} 
& T0014.001: Social Media Advertising & Deployed targeted Facebook ads focusing on high-income demographics & D0002.001: New accounts with high ad spending, suspicious targeting \\
\cline{2-4}
& T0017.001: Exploiting Greed & Emphasized ``limited-time opportunities'' and ``insider information'' & D0001.012: Psychological manipulation tactics, urgency-based language \\
\cline{2-4}
& T0020.003: Impersonating Celebrities & Used photos and names of renowned financial experts & D0001.002: Reverse image search reveals stolen content \\
\hline

\multirow{3}{2.3cm}{Engagement (P0003)} 
& T0021.001: Investment App Download & Required victims to download ``exclusive'' trading applications & D0003.002: Suspicious IP origins, unverified app sources \\
\cline{2-4}
& T0033: Fund Transfer Requests & Guided victims through wire transfers to offshore accounts & D0004.003: Unusual fund flow patterns, high-risk destination countries \\
\cline{2-4}
& T0034.002: Direct Fund Transfer & Executed actual monetary transfers from victim accounts & D0004.001: Transaction frequency anomalies, large sum movements \\
\hline

\multirow{3}{2.3cm}{Concealment (P0004)} 
& T0047.003: Shell Company Usage & Utilized multiple shell companies to obscure fund trails & D0004.007: Account lifecycle monitoring, entity relationship analysis \\
\cline{2-4}
& T0050.002: Domain Deletion & Removed fraudulent websites and domains post-operation & D0003.001: Domain lifecycle tracking, rapid takedown patterns \\
\cline{2-4}
& T0056: Cryptocurrency Usage & Converted funds to cryptocurrency for enhanced anonymity & D0004.008: Blockchain analysis, mixer service detection \\
\hline
\end{tabularx}
\label{tab:fist_case_mapping}
\end{table}


Table~\ref{tab:fist_case_mapping} illustrates the modularity of FIST and how detection opportunities arise at each operational phase. By identifying behavioral and technical signals, security teams can deploy targeted controls and improve incident response effectiveness.


\subsection{FIST Mapping (Highlights)}
\begin{itemize}
    \item \textbf{Preparation}: Collected victim profiles, fabricated credible investment platforms.
    \item \textbf{Promotion}: Deployed targeted messages and deepfake expert personas.
    \item \textbf{Engagement}: Built trust via ongoing interactions and fake performance metrics.
    \item \textbf{Concealment}: Used money mules, fragmented transactions, deleted records.
\end{itemize}


\section{Applications and Discussion}

FIST supports:
\begin{itemize}
    \item FIST enables the establishment of unified analysis standards, facilitating collaboration, communication, and knowledge accumulation across different organizations.
    \item Through structured mapping and technical indicators, it enhances the real-time identification of fraudulent anomalies in automated monitoring systems.
    \item The framework also supports the development of scalable fraud databases and threat intelligence sharing platforms.
    \item Additionally, FIST can be used to design training courses and educational materials, raising public and societal awareness of fraud prevention.
\end{itemize}


The FIST framework also establishes a foundation for describing fraud incidents, which facilitates advanced AI-driven behavior analytics, knowledge graph construction, and automated red-teaming and blue-teaming exercises. By providing a consistent taxonomy and structured dataset, FIST enables both technical automation and expert-driven threat hunting.


\section{Conclusion and Future Work}

The FIST framework offers a unified, extensible approach to fraud threat modeling, integrating technical, behavioral, and operational domains. By structuring fraud incidents into phases, tactics, and techniques, FIST enables detailed analysis, supports automated detection, and facilitates consistent intelligence sharing across organizations. The modular design allows for flexible adaptation to emerging threats and diverse fraud scenarios, bridging gaps between cybersecurity, criminology, and behavioral research.


In future work, we will further expand the technique and indicator catalog, incorporate advanced detection logic, and strengthen interoperability with existing security frameworks such as MITRE ATT\&CK and DISARM. We also plan to develop comprehensive case studies and empirical validations in collaboration with industry, government, and academic partners, aiming to assess FIST's effectiveness in real-world environments. Additionally, we will explore integrating FIST with AI-driven behavior analytics, knowledge graphs, and automated incident response systems to enhance both technical automation and expert-led threat hunting.


FIST is released as an open-source project, and we invite contributions from the community to continuously refine the framework, expand its knowledge base, and share best practices. Through collaborative development and transparent knowledge sharing, we hope FIST will accelerate progress in anti-fraud research, operations, and cross-sector defense capabilities.


\section*{Ethical Considerations}

All analyses in this study are conducted exclusively using publicly available or anonymized data, with no collection or use of personally identifiable information. The FIST framework is intended solely for the advancement of fraud detection, prevention, and defense, and is not designed or authorized for any offensive or malicious purposes. No human subjects were directly involved at any stage of this research, and all case studies are based on simulated or publicly documented incidents. The authors are committed to upholding the highest ethical standards and ensuring that the framework is applied responsibly to promote public safety and the common good.


\bibliographystyle{unsrt}

\begin{thebibliography}{1}
\bibitem{mitre_attack}
MITRE Corporation. ``ATT\&CK: Adversarial Tactics, Techniques, and Common Knowledge,'' 2024. \url{https://attack.mitre.org/}

\bibitem{disarm}
DISARM Foundation C.I.C. ``DISARM Framework,'' 2024. \url{https://disarm.foundation/}

\bibitem{sarkar2023}
G.~Sarkar, S.~Chaudhary, and R.~Goyal, ``Tactics, Techniques and Procedures of Cybercrime: A Methodology and Tool for Cybercrime Investigation Process,'' in \textit{Proc. 18th Int. Conf. Availability, Reliability and Security (ARES)}, Benevento, Italy, Aug. 2023, pp.~1--10.

\bibitem{iitkanpur2023}
PIB Delhi, ``Cybercrime Investigation Tool developed can track cyberattacks targeting human,'' Press Information Bureau, Govt. of India, Sep. 2023. \url{https://pib.gov.in/PressReleaseIframePage.aspx?PRID=1956941}

\bibitem{alhashmi2024}
A.~A.~Alhashmi, F.~Alzahrani, and M.~Alghamdi, ``Decoding the Deception: A Comprehensive Analysis of Cyber Scam Vulnerability Factors,'' \textit{Journal of Intelligent Systems and Applied Data Science}, vol.~2, no.~1, pp.~29--41, Apr. 2024.

\bibitem{stix}
OASIS. ``STIX (Structured Threat Information Expression),'' 2024. \url{https://oasis-open.github.io/cti-documentation/}
\end{thebibliography}

\end{document}